\documentclass[a4paper,draft]{amsart}

\usepackage{amsmath}
\usepackage{amssymb}
\usepackage{layaureo}

\usepackage[english,francais]{babel}

\newtheorem{theorem}{Theorem}[section]

\newtheorem{e-proposition}[theorem]{Proposition}

\newtheorem{definition}[theorem]{Definition\rm}

\newtheorem{irrider}{Definition}[section]
\newtheorem{defKdV}[irrider]{Definition}

\theoremstyle{plain}

\def\og{\leavevmode\raise.3ex\hbox{$\scriptscriptstyle\langle\!\langle$~}}
\def\fg{\leavevmode\raise.3ex\hbox{~$\!\scriptscriptstyle\,\rangle\!\rangle$}}

\DeclareMathOperator{\fder}{\mathcal{D}}
\DeclareMathOperator{\DD}{D}
\DeclareMathOperator{\e}{e} 
\newcommand{\toder}[3][]{\frac{{\mathrm d^{#1}}#2}{{\mathrm d #3}^{#1}}}
\newcommand{\parder}[3][]{\frac{{\partial^{#1}}#2}{{\partial #3}^{#1}}}
\newcommand{\rrfder}[3][]{{}_{#2}\partial_{#3}^{#1}}

\newcommand{\RRfder}[3][]{\sideset{_{#2}}{_{#3}^{#1}}\fder}

\begin{document}
% place in the next line the header (rubrique) chosen for your article,
% if you know it (you can also have 2, format : Header1/Header2
\selectlanguage{english}
\title{A Theorem on the Existence of Symmetries of Fractional PDEs}

\selectlanguage{english}
\author{Rosario Antonio Leo}
\address{Dipartimento di Matematica e Fisica ``Ennio De Giorgi", Universit\`a del Salento, Via per Arnesano, 73100 -- Lecce, Italy}
\email{leora@le.infn.it}
\author{Gabriele Sicuro}
\address{Dipartimento di Fisica ``Enrico Fermi", Universit\`a di Pisa, Italy}
\email{gabriele.sicuro@for.unipi.it}
\author{Piergiulio Tempesta}
\address{Departamento de Fisica Teorica II, M\'etodos Matem\'aticos de la f\'isica, Universidad Complutense de Madrid, Ciudad Universitaria, 28040, Madrid, Spain}
\email{ptempest@ucm.es}

\maketitle
\begin{abstract}
\selectlanguage{english}
% Text of abstract in English
We propose a theorem that extends the classical Lie approach to the case of fractional partial differential equations (fPDEs) of the Riemann--Liouville type in (1+1) dimensions.
\end{abstract}
%\end{frontmatter}

% main text
\section{Introduction}

The aim of this paper is to establish a general approach for the determination of Lie symmetries for fractional differential equations (FDEs) in (1+1) dimensions. Since the works of Abel, Riemann, Liouville, etc. in the XIX century, \cite{samko} these equations have been largely investigated. Especially  in the last decade, there has been a resurgence of interest, due to their manifold applications in statistical mechanics, economics, social sciences and nonlinear phenomena like anomalous diffusion.

The main theorem proposed here, concerning the existence of symmetries for FDEs, generalizes the very few results known in the literature. In \cite{BL,gazizov2009}, the case of equations involving fractional derivatives with respect to one independent variable has been considered. In \cite{gorenflo2000}, \cite{luchko}, interesting scale invariant solutions of diffusion equations have been constructed. The intrinsic \textit{noncommutativity} of the fractional derivatives with respect to different variables, and - in the case of a single variable - with respect to different fractional orders, has represented until now the main problem in the treatment of symmetries of fractional PDEs.

Our strategy is inspired by the classical Lie theory: the annihilation of the prolonged action of the vector fields generating the symmetry transformations is imposed. This condition leads to a system of determining equations that allow to deduce the explicit expression for the symmetry generators. The knowledge of the invariants associated with such generators is a sufficient condition to reduce a given fractional partial differential equation into a new one, characterized by a smaller number of independent variables. In the case of a fractional ODE, the reduction process leads to another fractional ODE of reduced order.

In this paper, we shall focus on the case of the Riemann--Liouville fractional calculus.
%The results obtained in this paper pave the way to the creation of a complete theory of Lie symmetries for FDEs, that would include generalizations to higher order symmetries, nonclassical symmetries, master symmetries, etc.

%A crucial issue, still completely unexplored at least to our knowledge, is the problem of integrability of nonlinear fractional evolution equations. In this paper, we introduce a fractional version of the Korteweg--de Vries equation (FKdV), which possesses many of the symmetries of the standard KdV equation.

%\label{}
% etc, etc

% \appendix

% \section{}
% \label{}

% The Acknowledgements are an un-numbered section
%\section*{Acknowledgements}
% Acknowledgements text here
%\subsection{Riemann--Liouville fractional derivative}\label{rigorousRLdef}
%Let us define the \textit{Riemann--Liouville fractional derivative} and the \textit{Riemann--Liouville fractional partial derivative}.
%Let us start with the following
%\begin{definition}
Let $AC(\Omega)$ be the space of absolutely continuous functions on the interval $\Omega:=[a,b]\subset \mathbb{R}$. We denote by $AC^{n}(\Omega)$, $n\in\mathbb{N}$, the space of functions $f\colon\Omega\to\mathbb{R}$ such that $f\in C^{n-1}(\Omega)$ and $\toder[n-1]{f}{x}(x)\in {AC}(\Omega)$.
%\end{definition}
\begin{definition}
[Riemann--Liouville fractional operator] \label{R} Let $p\in\mathbb{R}^+$ and $f\colon[a,b]\subseteq\mathbb{R}\to\mathbb{R}$, with $f\in AC^{[p]+1}([a,b])$, $[p]\in\mathbb{N}_0:=\mathbb N\cup\{0\}$ such that $[p]\leq p<[p]+1$. Let $t\in(a,b)$. The \textit{Riemann--Liouville fractional integral of order $p$ and terminals $(a,t)$} is defined by
\begin{equation}\textstyle\RRfder[-p]{a}{t}f(t):=\frac{1}{\Gamma(p)}\int_a^t(t-\tau)^{p-1}f(\tau)\mathrm{d}\tau,\label{R2}
\end{equation}
whereas the \textit{Riemann--Liouville fractional derivative of order $p$ and terminals $(a,t)$} is defined by
\begin{equation}\textstyle\RRfder[p]{a}{t}f(t):=\toder[{[p]+1}]{}{t}\RRfder[{p-[p]-1}]{a}{t}f(t)=\frac{1}{\Gamma(1+[p]-p)}
\toder[{[p]+1}]{}{t}\int_a^t(t-\tau)^{[p]-p}f(\tau)\mathrm{d}\tau,\label{R1}
\end{equation}
where $\Gamma(x):=\int_0^\infty t^{x-1}\e^{-t}\mathrm{d} t$ is the Gamma function.
\end{definition}
When $p\equiv k\in\mathbb{N}$, the previous definitions coincide with the usual $k$th-fold integral ($\lim_{p\to k^\pm}\RRfder[-p]{a}{t}f(t)=\int_a^{t}\mathrm{d}\tau_1\int_a^{\tau_1}\mathrm{d}\tau_2\cdots\int_a^{\tau_{k-1}}\mathrm{d} \tau_k f(\tau_k)=\frac{1}{(k-1)!}\int_a^t(t-\tau)^{k-1}f(\tau)\mathrm{d}\tau$) and with the $k$th-order derivative respectively ($\lim_{p\to k^\pm}\RRfder[p]{a}{t}f(t)=\toder[k]{f(t)}{t}$). We can now define the \textit{partial Riemann--Liouville fractional derivative}. For the sake of simplicity, we will consider the case of a function $f$ of two variables $x_1$ and $x_2$.
\begin{definition}[Partial fractional derivative and total fractional derivative] Let $p\in\mathbb{R}^+$ and $f(x_1,x_2)\colon [a_1,b_1]\times[a_2,b_2]\to\mathbb{R}$, $[a_1,b_1]\subset\mathbb{R}$, $[a_2,b_2]\subset\mathbb{R}$, $f^{(k,0)}(x_1,x_2):=\frac{\partial^{k}f}{\partial x_1^k}(x_1,x_2)$ continuous and integrable $\forall k\in\mathbb{N}_0$ s.t.~$k\leq [p]+1$ and $\forall x_2\in[a_2,b_2]$. We define the \textit{partial Riemann--Liouville fractional derivative} by
\begin{equation}\textstyle
\rrfder[p]{a_1}{x_1}f(x_1,x_2)=\frac{1}{\Gamma(1+[p]-p)}\parder[{[p]+1}]{}{x_1}\int_{a_1}^{x_1}(x_1-t)^{[p]-p}f(t,x_2)\mathrm{d} t.
\end{equation}
Let $g\colon [a_1,b_1]\to [a_2,b_2]$ a function such that $f(x_1,g(x_1))$ satisfies the requirements of definition \ref{R} respect to the variable $x_1$. We define the total fractional derivative with respect to the variable $x_1$ by
\begin{equation}\textstyle
\RRfder[p]{a}{x_1}f(x_1,g(x_1)):=\frac{1}{\Gamma(1+[p]-p)}
\toder[{[p]+1}]{}{x_1}\int_a^{x_1}(x_1-\tau)^{[p]-p}f(\tau,g(\tau))\mathrm{d}\tau
\end{equation}
\end{definition}

%\begin{definition} Let $p,q\in\mathbb{R}^+$ and $f(x_1,x_2)\colon [a_1,b_1]\times[a_2,b_2]\to\mathbb{R}$, $[a_1,b_1]\subset\mathbb{R}$, $[a_2,b_2]\subset\mathbb{R}$, $f^{(k,l)}(x_1,x_2):=\frac{\partial^{k+l}f}{\partial x_1^k\partial x_2^l}(x_1,x_2)$ continuous and integrable $\forall k,l\in\mathbb{N}_0$ s.t.~$k\leq [p]+1$ and $l\leq [q]+1$. We define the mixed fractional derivative as
%\begin{equation}
%\RRfder[q]{a_2}{x_2}\RRfder[p]{a_1}{x_1}f(x_1,x_2)=\RRfder[q]{a_2}{x_2}\left(\RRfder[p]{a_1}{x_1}f(x_1,x_2)\right)
%\end{equation}
%\end{definition}
\noindent A mixed fractional derivative can be directly introduced. However, note that \[\rrfder[q]{a_2}{x_2}\rrfder[p]{a_1}{x_1}f(x_1,x_2)\neq\rrfder[p]{a_1}{x_1}\rrfder[q]{a_2}{x_2}f(x_1,x_2).\]
%Observe that, under the working hypotheses of the previous definition,
%\begin{multline}\textstyle\RRfder[p]{a_1}{x_1}\RRfder[q]{a_2}{x_2}f(x_1,x_2)=\frac{\partial^{[p]+[q]+2}}{\partial x_1^{[p]+1}\partial x_2^{[q]+1}}\int_{a_1}^{x_1}\int_{a_2}^{x_2}\frac{(x_1-t_1)^{[p]-p}(x_2-t_2)^{[p]-p}f(t_1,t_2)}{\Gamma(1+[p]-p)\Gamma(1+[q]-q)}\mathrm{d} t_1\mathrm{d} t_2\\=\RRfder[q]{a_2}{x_2}\RRfder[p]{a_1}{x_1}f(x_1,x_2).\end{multline}
%In fact, taking for simplicity $p,q\in(0,1)$, we have, by means of differentiation under the integral signs
%\begin{multline}
%\textstyle\RRfder[p]{a_1}{x_1}\RRfder[q]{a_2}{x_2}f(x_1,x_2)=\frac{1}{\Gamma(1-p)\Gamma(1-q)}\left[\frac{f(a_1,a_2)}{(x_1-a_1)^p(x_2-a_2)^q}\right.\\
%\textstyle+\frac{1}{(x_1-a_1)^p}\int_{a_2}^{x_2}\parder{f}{t_2}(a_1,t_2)\frac{\dd t_2}{(x_2-t_2)^q}+\frac{1}{(x_2-a_2)^q}\int_{a_1}^{x_1}\parder{f}{t_1}(t_1,a_2)\frac{\dd t_1}{(x_1-t_1)^p}\\
%\textstyle\left.+\int_{a_1}^{x_1}\int_{x_2}^{t_2}\frac{\partial^2f(t_1,t_2)}{\partial t_1\partial t_2}\frac{\dd t_1\dd t_2}{(x_1-t_1)^p(x_2-t_2)^q}\right]=\RRfder[q]{a_2}{x_2}\RRfder[p]{a_1}{x_1}f(x_1,x_2).
%\end{multline}
\section{Lie Theory for Fractional Partial Differential Equations}
Let us consider the case of \textsc{fpde}s (\textit{fractional partial differential equations}) with one dependent variable $u\in U\subseteq\mathbb{R}$ and two independent variables $(x_1,x_2)\in X\subseteq \mathbb{R}^2$. We suppose that the equation takes the form
\begin{equation}
\mathcal E\left(x_1,x_2,u,\rrfder[p_1,q_1]{a}{m(1),3-m(1)} u,\dots,\rrfder[p_K,q_K]{a}{m(K),3-m(K)} u\right)=0.\label{fPDE}
\end{equation}
Here $\mathcal E$ is a polynomial involving $K$ fractional derivatives of the form $\label{pq}\rrfder[p,q]{a}{1,2}u(x_1,x_2):=\rrfder[p]{a}{x_1}\rrfder[q]{a}{x_2} u(x_1,x_2)$ or $\rrfder[p,q]{a}{2,1}u(x_1,x_2):=\rrfder[p]{a}{x_2}\rrfder[q]{a}{x_1} u(x_1,x_2)$, where $K\in\mathbb N$, $m(i)\colon\{1,\dots,K\}\subset\mathbb N\to\{1,2\}$, $a\in\mathbb{R}$, and $p,q\in[0,+\infty)$ are not both zero. In the subsequent considerations, we shall assume that all fractional derivatives appearing in $\mathcal E$ have \textit{the same lower extreme} $a$.
A \textit{continuous symmetry group $G$} or \textit{Lie symmetry} for the equation $\mathcal E=0$ is a one-parameter group of continuous transformations that maps solutions $(x_1,x_2,u)\in X\times U:=\mathcal M$ into solutions $g\cdot(x_1,x_2,u)=(\tilde x_1,\tilde x_2,\tilde u)=(\Xi_g(x_1,x_2,u),\Phi_g(x_1,x_2,u))\in \mathcal M$, $g\in G$ for some functions $\Xi_g\colon \mathcal M\to X$, $\Phi_g\colon \mathcal M\to U$. A generic element $g\in G$ has the form $g=\e^{\epsilon\mathbf v}$, where $\epsilon\in\mathbb{R}$ is the parameter of the group transformation and $\mathbf v$ is a vector field generating $G$. We shall restrict to vector fields of the form
$\mathbf v=\xi^1(x_1,x_2,u)\partial_{x_1}+ \xi^2(x_1,x_2,u)\partial_{x_2} +\phi(x_1,x_2,u)\partial_u.$, i.e. we will study \textit{point symmetries}. We also assume that the action of the symmetry group $G$, $(x_1,x_2,u)\xrightarrow{g}(\tilde x_1,\tilde x_2,\tilde u)$, can be expressed by means of smooth functions such that $\left.\toder{\tilde x_i}{\epsilon}\right|_{\epsilon=0}=\xi^i(x_1,x_2,u)$, $i=1,2$, $\left.\toder{\tilde u}{\epsilon}\right|_{\epsilon=0}=\phi(x_1,x_2,u)$.
As in the case of standard differential equations \cite{olver}, we prolong the vector field as
\begin{multline}
\mathrm{pr}^{(\mathcal{E})}\mathbf{v}=\xi^1(x_1,x_2,u)\partial_{x_1}+ \xi^2(x_1,x_2,u)\partial_{x_2}+\phi(x_1,x_2,u)\partial_u+\\
+\sum_{\substack{l,m\in\mathbb N_0\\(l,n)\neq (0,0)}}\phi^{l,n}_{1,2}(x_1,x_2,u,\dots)\partial_{\partial^{l,n}_{1,2} u}\\+\sum_i{\sum_{\substack{k,r\in\mathbb N_0\\k-p_i\not\in\mathbb N,\ r-q_i\not\in\mathbb N}}}\phi^{(p_i-k,q_i-r)}_{{m(i)},{3-m(i)}}(x_1,x_2,u,\dots)\partial_{{}_a\partial^{p_i-k,q_i-r}_{{m(i)},{3-m(i)}}u}, \label{prolong}
\end{multline}
where $m(i)=1,2$, the sum $\sum_i$ runs over all the ordered couples of parameters $(p_i,q_i)$ such that at least one of the parameters selected among $p_i$ and $q_i$ is a non integer positive real number and $\rrfder[p_i,q_i]{a}{{m(i)},{3-m(i)}}u$ does appear in $\mathcal E$. By definition \[\phi^{(p,q)}_{m,{3-m}}:=\left.\frac{d}{d\epsilon}\left[\rrfder[p]{a}{\tilde x_m}\rrfder[q]{a}{\tilde x_{3-m}}\tilde u(\tilde x_1,\tilde x_2)\right]\right|_{\epsilon=0},\quad m=1,2.\]
%We look for an explicit expression for this quantity.
%where we have denoted by $\RRfder[P]{a}{(x,t)}u$ a fractional derivative in the irreducible form
%\begin{equation}
%\RRfder[P]{a}{(x,t)}u(x):= \RRfder[b_1,\dots,b_L]{a}{{k_1},\cdots,{k_L}}u(x),\qquad k_i=1,2;
%\end{equation}
%if $k_i=1$ the $i$th derivation is with respect to $x$, otherwise it is respect to $t$. The sum $\sum_P$ runs over all (infinite) possible terms satisfying the properties \ref{conditions}. Note that if all derivatives in $\mathcal E$ have integer order, the sum is finite as consequence of the first condition.

%We look for an explicit expression for $\phi_{P}$ as a function of the coefficients of the vector field $\textbf{v}$. We will denote by
%\begin{equation}
%\phi^{p_1,\dots,p_N}_{{k_1},\dots,{k_N}}:=\left.\frac{d}{d\epsilon}\left[\RRfder[p_1,\dots,p_N]{a}{\tilde x_{k_1},\dots,\tilde x_{k_N}}\tilde u(\tilde x)\right]\right|_{\epsilon=0}
%\end{equation}
%\section{Symmetries of Partial Fractional Differential Equations: the case 1+1 dimensional}
The following theorems represent the main results of the paper.
\begin{theorem}[Prolongation formula]  Assume that $G$ is a local group of transformations acting on $\mathcal M=X\times U$. Then for $m=1,2$ and $p,q\in(0,+\infty)$, we have the following explicit expressions for the coefficients of the prolonged vector field \eqref{prolong}:
\begin{subequations}
\begin{multline}\label{phi0pa}\phi^p_m=\RRfder[p]{a}{m}\phi+\RRfder[p]{a}{m}\left(u\DD_m\xi^m\right)-\RRfder[p+1]{a}{m}\left(\xi^m u\right)\\+\xi^m\RRfder[p+1]{a}{m}u+\xi^{3-m}\RRfder[p]{a}{m}\partial_{3-m} u-\RRfder[p]{a}{m}\left(\xi\partial_{3-m}u\right)\end{multline}
\begin{multline}\label{genprolongation}
\phi^{p,q}_{m,3-m}=\RRfder[p,q]{a}{m,3-m}\left(\varphi-\sum_{i=1}^2\xi^i \partial_{i} u\right)+\sum_{i=1}^2\xi^i\partial_{i}\rrfder[p,q]{a}{m,3-m}u+\RRfder[p,q]{a}{m,3-m}\DD_{3-m}\left(\xi^{3-m} u\right),\\
+\RRfder[p]{a}{m}\DD_m\left(\xi^m \rrfder[q]{a}{3-m}u\right)-\RRfder[p,q+1]{a}{m,3-m}\left(\xi^{3-m} u\right)-\RRfder[p+1]{a}{m}\left(\xi^m \rrfder[q]{a}{3-m}u\right),
\end{multline}
where we use the notations $\partial_i:=\frac{\partial}{\partial x_i}$ and $\DD_i:=\frac{\DD}{\DD x_i}$ for the partial and total derivative respectively. In particular, by taking $a=0$, we have \cite{gazizov2009}
\begin{multline}\label{phi0p}\phi^p_m=\rrfder[p]{0}{m}\phi+\RRfder[p]{0}{m}u\left(\partial_u\phi-p\DD_m\xi^m\right)\\+\sum\limits_{n=2}^\infty \sum\limits_{l=2}^n\sum\limits_{k=2}^l\sum\limits_{r=0}^{k-1}\binom{p}{n}\binom{n}{l}\binom{k}{r}\frac{x_m^{n-p}(-u)^r}{k!\Gamma(n+1-q)}\frac{d^l}{dx_m^l}\left(u^{k-r}\right)\frac{\partial^{n-l+k}\phi}{\partial x_m^{n-l}\partial u^k}-u\,\rrfder[p]{0}{m}\partial_u\phi
\\+\sum\limits_{n=1}^\infty\left\{\left[\binom{p}{n}\partial^n_m\partial_u\phi-\binom{p}{n+1}\DD^{n+1}_m\xi^m\right]\RRfder[p-n]{0}{m}u-\binom{p}{n}\DD_m^n\xi^{3-m}\partial_{3-m}\RRfder[p-n]{0}{m}u\right\}.
\end{multline}\end{subequations}
\end{theorem}
\begin{theorem}[Symmetries for FPDEs, case 1+1] Under the hypotheses of the previous theorem, given a \textsc{fpde} of the form \eqref{fPDE}, if the relation
\begin{equation}
\left.\mathrm{pr}^{(\mathcal{E})}\mathbf v (\mathcal{E})\right|_{\mathcal{E}=0}=0,
\end{equation}
holds, then $ \mathbf v$ is the generator of a Lie symmetry of eq. \eqref{fPDE}.
\end{theorem}

%\paragraph*{\textit{Sketch of the Proof.}}
%For the derivation of \eqref{phi0pa} and \eqref{phi0p} we refer to \cite{gazizov2009}. Taking $p,q\neq 0$, to prove the eq.~\eqref{genprolongation}, we observe that $\phi^{p,q}_{m,3-m}$, $m=1,2$, can be %obtained by a prolongation procedure using \eqref{phi0pa} on a vector field prolonged up to $\RRfder[q]{a}{3-m}u$, $m=1,2$:
%\begin{multline}
%\phi^{p,q}_{m,3-m}=\RRfder[p]{a}{m}\left(\phi^{q}_{3-m}-\sum_{i=1}^2\xi^i\partial_{i}\RRfder[q]{a}{3-m} %u\right)+\sum_{i}\xi^i\partial_{i}\RRfder[p]{a}{m}\RRfder[q]{a}{3-m}u\\+\RRfder[p]{a}{m}\DD_{m}(\xi^{m}\RRfder[q]{a}{3-m} u)-\RRfder[p+1]{a}{m}(\xi^{m}\RRfder[q]{a}{3-m} u).
%\end{multline}
%Expressing $\phi^q_{3-m}$ using eq.~\eqref{phi0p} we complete the proof. Note that if $p,q\in\mathbb N$ we recover the classical prolongation formula.\qed

%\subsection{Example of symmetry reduction.}
As an application of the previous theory, we propose a symmetry analysis of a fractional KdV--Burgers equation.
\begin{defKdV}[Fractional KdV--Burgers equation]
We shall call the equation
\begin{equation}
\rrfder[p]{0}{x_2}u+u\,\rrfder[q]{0}{x_1}u+\rrfder[r]{0}{x_1}u=0,\quad p,q,r\in\mathbb{R}^+.\label{frKdV}
\end{equation}
the fractional Korteweg--de Vries--Burgers equation.
\end{defKdV} This form of the \textsc{fKdV--Burgers} equation, at the best of our knowledge, is new. In the following, we will adopt the short notation $u^{(p,0)}:=\rrfder[p]{0}{x_1}u$ and $u^{(0,q)}:=\rrfder[q]{0}{x_2}u$.
%\subsection{General fractional case: $p,q,r\in\mathbb{R}^+_0\setminus\mathbb{N}$}
We consider the case $p,q,r\in\mathbb{R}/\mathbb{Z}$,  with $q<r$. The determining equation takes the form
\begin{equation}
\left.\left(\phi^p_2+u\phi^q_1+\phi {u}^{(q,0)}+\phi^r_1\right)\right|_{{u}^{(0,p)}+u{u}^{(q,0)}+{u}^{(r,0)}=0}=0.
\end{equation}
We no longer have a translational symmetry. We get uniquely the symmetry generator of scaling transformations
\begin{equation}\label{vfrKdV}
\mathbf v=px_1\partial_1+rx_2\partial_2+p(q-r)u\partial_u.
\end{equation}
If $q=r$, it can be similarly proved that the symmetry generator is $\mathbf v=px_1\partial_1+rx_2\partial_2$. If $q\neq r$ we can obtain an invariant by means of the relation $\mathbf v\eta(x_1,x_2,u)=0$. %Using the method of characteristics, we have that $x_1x_2^{-{p\over r}}$ and $x_2^{-{1\over r}} u^{1\over p(q-r)}$ are invariant quantities under the action of the transformations generated by $\mathbf v$.
Consequently, we perform a symmetry reduction by looking for a solution of the form $u(x_1,x_2)=x_2^{p(q-r)\over r} v(x_1x_2^{-{p\over r}}). \label{sol}$
%To perform the reduction, suppose that we have a function $\varphi(x,t)\equiv\varphi(xt^{-\alpha})$ and and denote by $z=xt^{-\alpha}$, $\alpha>0$.
It is easily proved that
\begin{equation}
\rrfder[p]{0}{x_2}v(x_1 x_2^{-\alpha})=x_{2}^{-p}\left(\mathfrak{D}^{1-p,p}_{1\over\alpha}v\right)(x_1 x_2^{-\alpha}), \hspace{10mm} \alpha=p/r
\end{equation}
where $\left(\mathfrak{D}^{c,a}_{b}f\right)(y):= \prod_{j=0}^{[a]}\left(j+c- {y\over b}\toder{}{y}\right)\mathfrak K^{c+a,[a]+1-a}_{b}f(y)$ is the \textit{Erd\'ely--Kober fractional differential operator of order $a\geq 0$} and $
\left(\mathfrak K^{c,a}_b f\right)(y):=\frac{1}{\Gamma(a)} \int_1^\infty(\eta-1)^{a-1}\eta^{-(a+c)}f(y\eta^{1\over b})d\eta$, $a>0$, $b,c\in\mathbb R$, is the \textit{Erd\'elyi--Kober fractional integral operator}.
Using the relation $\rrfder[p]{0}{x}f(\lambda x)=\lambda^p\rrfder[p]{0}{\lambda x}f(\lambda x)$, one gets the reduced equation in the form:
\begin{equation}
z^{q-r}\mathfrak{D}^{1-p,p}_{r\over p}\left(z^{r-q}v(z)\right)+v(z)\RRfder[q]{0}{z}v(z)+\RRfder[r]{0}{z}v(z)=0,\hspace{10mm} z=x_1x_2^{-{p\over r}}.\label{reducedKdV}
\end{equation}
%\begin{equation}u(x,t)=\frac{a}{\chi}\left(1+\frac{\chi}{\sqrt{\pi}}\frac{t}{\sqrt{x}}\right).\label{4221}\end{equation}
This equation can be solved numerically. Its solutions, by means of \eqref{sol}, will provide invariant solutions of the KdV--Burgers equation.
\section*{Acknowledgment}
G.~S. wish to thank Dr. Alexei Kasatkin for useful discussions. The research of P.~T. has been partly supported by the grant FIS2011--22566, Ministerio de Ciencia e Innovaci\'{o}n, Spain.

\end{document}